\documentclass{iau_FM}
\usepackage{graphicx}

\title[FM12.~~Interstellar dust modelling] 
{Interstellar dust modelling: Interfacing laboratory, theoretical and observational studies (The THEMIS model)}

\author[Anthony P. Jones]   
{Anthony P. Jones}

\affiliation{Institut dÕAstrophysique Spatiale, UMR8617, CNRS/Universit\'e Paris Sud, Universit\'e Paris-Saclay, Universit\'e Paris Sud, Orsay F-91405, France \\email: {\tt Anthony.Jones@ias.u-psud.fr}}

\pubyear{2015}
\jname{Astronomy in Focus, Volume 1} 
\editors{Piero Benvenuti, ed.}
\begin{document}

\maketitle

\begin{abstract}
The construction of viable and physically-realistic interstellar dust models is only possible if the constraints imposed by laboratory data on interstellar dust analogue materials are respected and used within a meaningful theoretical framework. These ``physical'' dust models can then be directly compared to observations without the need for any tuning to fit the observations. Such models will generally fail to achieve the excellent fits to observations that ``empirical'' models are able to achieve. However, the physically-realistic  approach will necessarily lead to a deeper insight and a fuller understanding of the nature and evolution of interstellar dust.
The THEMIS modelling approach, based on (hydrogenated) amorphous carbons and amorphous silicates with metallic Fe and/or FeS nano-inclusions appears to be a promising move in this direction. 
\end{abstract}

\firstsection 
\section{Introduction}

In order for a dust model to be viable it must be consistent with as wide a possible range of dust observables ({\it e.g.}, pre-solar grain compositions, elemental abundances/depletions, extinction, absorption, scattering, emission, infrared spectra, polarisation, x-ray absorption and scattering, \ldots) and the variations of those observables across the interstellar medium (ISM). However, each of these observables is necessarily selective because none of them are un-biased. For example, the analysed pre-solar grains ({\it e.g.}, \cite[Anders \& Zinner 1993]{AndersZinner93}) and the analysis of the STARDUST interstellar grains, (\cite[Westphal, Stroud, Bechtel, et al. 2014]{Westphal14}) represent incomplete and/or selective samplings. Additionally, interstellar dust observations have not yet fully-sampled all dust in all environments.  

As a minimum, but insufficient, requirement the optical properties of all dust components must be consistent with the Kramers-Kronig relations, {\it i.e.}, the real ($n$) and imaginary ($k$) parts of the complex index of refraction ($m=n+ik$) must be self-consistent over as wide as possible a range of wavelengths. However, just because this condition is fulfilled does not necessarily imply that the material is physically-realisable, {\it i.e.}, that it can actually exist. The only valid test of the physical-reasonableness of an interstellar dust analogue material is that it, or a very closely-related material, can be synthesised, characterised and analysed in the laboratory. Hence, interstellar dust models must, wherever possible, be guided by laboratory measurements of physically-realised materials.

\section{More complex but more realistic dust modelling (THEMIS)}

It is clear that the dust optical properties depend upon the material composition and structure. Some materials can show a wide range of structures for a given chemical composition and all optical properties depend upon the particle size below some limiting dimension. Amorphous carbons, a-C, and hydrogenated amorphous carbons, a-C:H, are a prime example of this. These materials, under the collective term a-C(:H) exhibit wide-ranging properties, from insulators to conductors (but are mostly semiconductors), from hydrogen-rich to hydrogen-poor and from highly absorbing to highly scattering. This impressive gamut of properties from the large family of a-C(:H) materials is rather impressive for a material consisting of only carbon and hydrogen atoms. The suite of a-C(:H) materials is now receiving some attention within the astrophysical community.  

The structures, compositions and size-dependent optical properties for a-C(:H) materials were recently derived for a wide range of H/C ratios and particles sizes over a broad wavelength range (soft x-ray to cm, the optEC$_{(s)}$ and optEC$_{(s)}$(a) datasets, \cite[Jones 2012a,b,c]{Jones12a}). These data were built ground up from theoretical considerations and strongly-constrained by laboratory data. The application of these data within the framework of a diffuse ISM dust model (\cite[Jones et al. 2013] {Jones13}; \cite[K\"ohler, Jones \& Ysard 2014]{Koehler14}) appears to be consistent with most of the dust observables ({\it e.g.}, see Fig.\,\ref{fig1}). This model comprises log-normal size distributions of large ($a \sim 150$\,nm) core/mantle grains of amorphous silicate and a-C:H cores with a-C mantles, and  a-C nano-particles with with a power-law size distribution biased towards the smallest grains. The model satisfies the physically-reasonable requirement in the sense that it uses only optical properties that are either directly measured in the laboratory or that are firmly-anchored to laboratory measurements and that are consistent with detailed theoretical considerations (\cite[Jones 2012a,b,c]{Jones12a}). The \cite[Jones et al. (2013)]{Jones13}/\cite[K\"ohler, Jones \& Ysard (2014)]{Koehler14} dust model uses the laboratory-constrained  a-C(:H) and olivine-type (a-Sil$_{\rm ol}$) and pyroxene-type (a-Sil$_{\rm px}$) amorphous silicate data {\it as is} without any tuning to match specific astronomical observations. The result is a model that matches reasonably well most of the interstellar dust observables without the need to ``tweak" the input data. However, the model has evolved, much as dust does in the ISM, to encompass the changes in the dust optical and physical properties as it interacts with and responds to its immediate environment. We now refer to our standard dust model, all of the associated dust evolution studies and future developments and extensions of this work under the umbrella acronym THEMIS (The Heterogeneous dust Evolution Model at the IaS). THEMIS is a key part of the European FP7 DustPedia project (DustPedia.com). The following summarises key elements of our modelling work: 

\begin{figure}[b]
\begin{center}
 \includegraphics[width=3.4in]{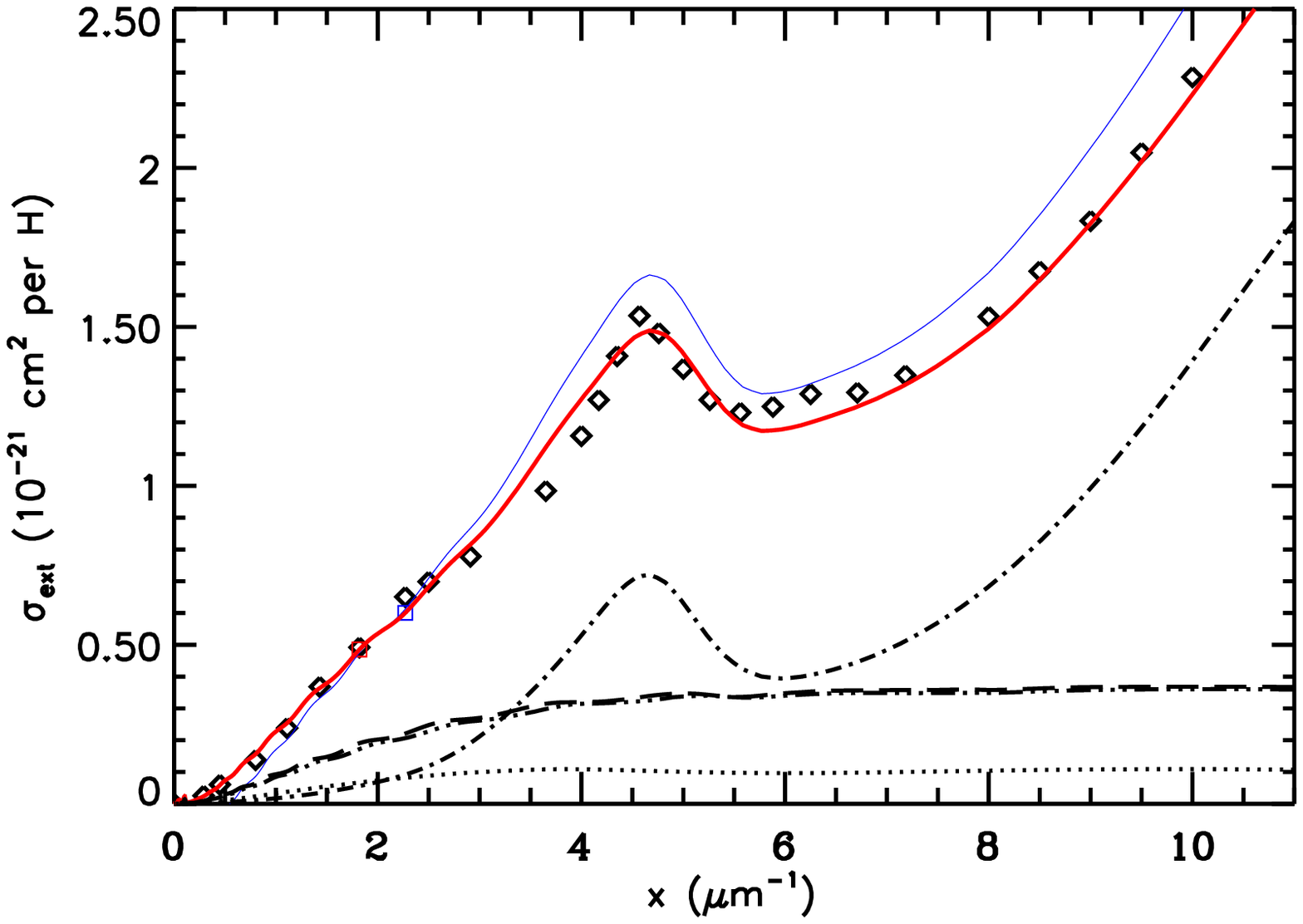} 
 \includegraphics[width=3.4in]{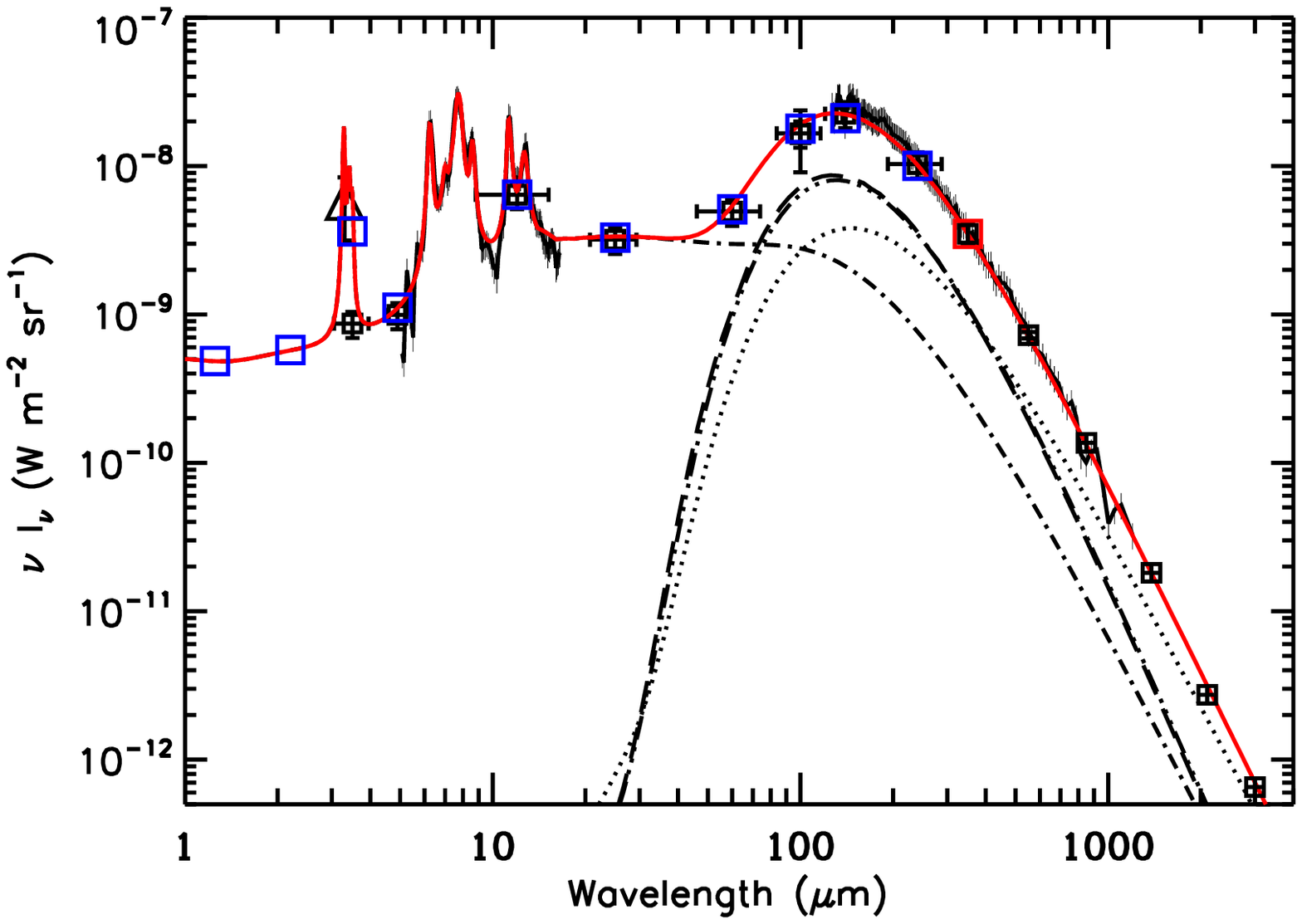}  \caption{The \cite[Jones et al. (2013)]{Jones13}/\cite[K\"ohler, Jones \& Ysard (2014)]{Koehler14} diffuse ISM dust model visible-UV extinction cross-sections (top) and the dust IR-mm SED (for $N_{\rm H} = 1 \times 10^{20}$\,cm$^{-2}$, bottom). The red lines show the totals in each case, comprising contributions from  a-C coated a-C:H (dotted), a-Sil$_{\rm ol}$ (long-dashed) and a-Sil$_{\rm px}$ (triple dot-dashed) grains, and small a-C grains (dot-dashed). The blue line in the upper plot shows the B and V band-normalised cross-section data.}
   \label{fig1}
\end{center}
\end{figure}

{\underline{\it Dust evolution}}. Dust is clearly not the same everywhere and significant differences are observed within and between given phases of the ISM. The THEMIS dust modelling approach has the in-built capacity to account for dust differences in a physically-meaningful way through variations in the dust optical properties, size distributions and structure as a function of the local ISM conditions, principally the interstellar radiation field and the gas density ({\it e.g.}, \cite[Jones 2012a,b,c]{Jones12a}, \cite[Jones et al. 2013]{Jones13}). 

{\underline{\it Dust in the diffuse ISM}}. The \cite[Jones et al. (2013)]{Jones13}/\cite[K\"ohler, Jones \& Ysard (2014)]{Koehler14} dust model was developed to match the dust properties in the diffuse ISM and comparisons with recent Planck observations of dust in these regions appear to indicate that the model is consistent with the observed variations in the dust properties (\cite[Ysard et al. 2015a]{Ysard15a}; \cite[Fanciullo et al. 2015]{Fanciullo15}). The principal processes encompassed by our model of large grain  evolution in the ISM are variations in the mantling materials ({\it i.e.}, depth and  composition) and in the relative masses of the amorphous silicate and carbonaceous dust components (\cite[Ysard et al. 2015a]{Ysard15a}).

{\underline{\it Dust in low energy, high density regions}}. The model has also been extended to follow the evolution of the dust properties in the transition between diffuse and dense interstellar media (\cite[Jones et al. 2014]{Jonesetal14}; \cite[K\"ohler, Ysard \& Jones 2015]{Koehler15}), including a self-consistent explanation of both cloud-shine and core-shine (\cite[Jones et al. 2015]{Jonesetal15}; \cite[Ysard et al. 2015b]{Ysard15b}). In these environments grain growth through a-C(:H) mantle formation and grain-grain coagulation are the fundamental actors. 

{\underline{\it Dust in high energy, low density regions}}. The a-C(:H) properties also seem to provide viable routes to interstellar/circumstellar fullerene formation in planetary nebul\ae\ (\cite[Micelotta et al. 2012]{Micelotta12}) and to the formation of molecular hydrogen in moderately excited photo-dissociation regions (PDRs, \cite[Jones \& Habart 2015]{JonesHabart15}). Further, dust evolution in energetic environments such as supernova-driven shock waves and a hot coronal gas has also been explored with the new dust model (\cite[Bocchio et al. 2012]{Bocchio12}; \cite[Bocchio et al. 2013]{Bocchio13}; \cite[Bocchio, Jones \& Slavin 2014]{Bocchio14}). This work is now being extended to a consideration of dust evolution in H\footnotesize{II} regions, galactic haloes and the intergalactic medium (IGM). Erosive processes involving high energy photons, electrons and ions are the major dust modifiers and destructors in the energetic regions of the ISM and the IGM. 

The THEMIS approach provides a detailed framework within which the solutions to several interesting interstellar conundrums such as ÒvolatileÓ silicon in PDRs, sulphur and nitrogen depletions, the origin of the blue and red photoluminescence and the diffuse interstellar bands (DIBs) may yet be found (\cite[Jones 2013]{Jones13}; \cite[Jones 2014]{Jones14}).

\section{Conclusions}

Interstellar dust is clearly not the same everywhere, it evolves in response to its immediate environment. For example in the transition from diffuse to denser regions of the ISM the dust will accrete a-C(:H) and ice mantles and coagulate to form aggregate grains with increasing density. Empirically-based models, which lack the necessary grounding in laboratory data, can never capture or fully uncover the physical origins of dust evolution even though they may provide excellent fits to the observational data. On the contrary, interstellar dust models strictly-anchored to laboratory measurements on physically-realisable and physically-realised dust analogue materials can provide a much deeper physical insight, even when the fit to the observational data is less that `perfect'. The aim of such modelling should indeed not be a perfect fit to the observational data but the best-possible re-construction of those data with an astute use of measurement-constrained dust properties. 

\vspace*{0.5cm}\noindent {\bf Acknowledgements:} I would like to thank my collaborators, Melanie K\"ohler, Nathalie Ysard, Marco Bocchio, Aur\'elie R\'emy-Ruyer, Elisabetta Micelotta, Lapo Fanciullo, Laurent Verstraete, Vincent Guillet, Emilie Habart, Alain Abergel, Emmanuel Dartois, Lisseth Gavilan and Marie Godard for innumerable hours of discussion about interstellar dust in all its guises.

\end{document}